\documentclass[12pt,preprint]{aastex}

\shorttitle{Fe XI lines in SERTS active region spectrum}
\shortauthors{F. P. Keenan et al.}
\begin{document}

\title{Soft X-ray emission lines of \ion{Fe}{15} in solar flare observations and
the Chandra spectrum of Capella}

\author{F. P. Keenan\altaffilmark{1}, J. J. Drake\altaffilmark{2},
S.~Chung\altaffilmark{2}, N.~S. Brickhouse\altaffilmark{2}, 
K. M. Aggarwal\altaffilmark{1}, 
A. Z. Msezane\altaffilmark{3},  R. S. I. Ryans\altaffilmark{1} 
and D. S. Bloomfield\altaffilmark{1}}

\email{F.Keenan@qub.ac.uk}

\altaffiltext{1}{Department of Physics \& Astronomy, Queen's University Belfast,
Belfast, BT7 1NN, Northern Ireland, U.K.}

\altaffiltext{2}{Smithsonian Astrophysical Observatory (SAO), MS 3,
60 Garden Street, Cambridge, MA 02138}

\altaffiltext{3}{Center for Theoretical Studies of Physical Systems,
Clark Atlanta University, Atlanta, GA 30304}



\begin{abstract}

Recent calculations of atomic data for \ion{Fe}{15} have been used to
generate theoretical line ratios
involving {\em n} = 3--4 transitions in the soft X-ray spectral region
($\sim$52--83 \AA), for a wide range of electron temperatures and
densities applicable to solar and stellar
coronal plasmas. A comparison of these
with solar flare observations from a rocket-borne spectrograph (XSST)
reveals generally good agreement between 
theory and experiment. In particular, 
the 82.76 \AA\ emission line in the XSST spectrum
is identified, for the first time to our knowledge in an
astrophysical source, as the 
3{\em s}3{\em d} $^{3}$D$_{3}$--3{\em s}4{\em p} $^{3}$P$_{2}$ transition of
\ion{Fe}{15}. 
Previous suggested 
identifications of the 53.11, 63.97 and 69.65 \AA\ features as the
3{\em s}$^{2}$ $^{1}$S--3{\em s}4{\em p} $^{3}$P$_{1}$,
3{\em p}$^{2}$ $^{1}$D--3{\em s}4{\em f} $^{1}$F
and
3{\em s}3{\em p} $^{1}$P--3{\em s}4{\em s} $^{1}$S lines of 
\ion{Fe}{15}, respectively, are confirmed. However the
former is blended (at the 
$\sim$ 50\%\ level) with the \ion{S}{9} 
2{\em s}$^{2}$2{\em p}$^{4}$ $^{3}$P$_{1}$--2{\em p}$^{3}$3{\em s} $^{3}$P$_{2}$
transition.
Most of the \ion{Fe}{15}
transitions which are blended have had the species
responsible clearly identified, although there 
remain a few instances where this has not been
possible.
The line ratio calculations are also compared 
with a co-added spectrum of Capella obtained
with the {\em Chandra} satellite, which is probably the highest
signal-to-noise observation achieved for a stellar source
in the $\sim$25--175 \AA\ soft X-ray region. Good agreement is found between
theory and experiment, indicating that the \ion{Fe}{15} 
lines are reliably detected 
in {\em Chandra} spectra, and hence may be employed
as diagnostics to determine the 
temperature and/or density of the emitting plasma. However the
line blending in the {\em Chandra} data is such that 
individual emission lines are difficult to measure 
accurately, and fluxes may only be reliably
determined via detailed profile fitting of the observations.
The co-added Capella spectrum is made available to
hopefully encourage further exploration of the soft X-ray region in
astronomical sources. 
\end{abstract}



\section{Introduction}

Emission lines arising from {\em n} = 3--3 transitions in \ion{Fe}{15}
are widely detected in solar extreme ultraviolet (EUV) spectra
covering the $\sim$200--400 \AA\ wavelength interval (Dere 1978;
Thomas \& Neupert 1994).  The 3{\em s}$^{2}$ $^{1}$S--3{\em s}3{\em p}
$^{1}$P resonance line at 284 \AA\ is one of the most intense emission
features in the solar EUV spectrum, and has also been widely observed
in stellar coronal sources by the {\em Extreme Ultraviolet Explorer}
(EUVE) satellite (Laming \& Drake 1999). Bely \& Blaha (1968) first
noted the potential of EUV lines of \ion{Fe}{15} as diagnostics for
the emitting plasma, and since then many authors have calculated
emission line intensities for this ion and compared these with solar
observations (see, for example, Kastner \& Bhatia 2001 and references
therein).

However \ion{Fe}{15} also shows a rich emission line spectrum in the
soft X-ray region, $\sim$50--80 \AA, arising from {\em n} = 3--4
transitions.  By contrast to the EUV lines, there has been little work
on the soft X-ray transitions in the astrophysical literature,
probably as a result of the limited availability of high quality solar
spectra for this wavelength region.  There were several rocket-borne
flights in the 1960's which covered the soft X-ray region, but these
detected few or no \ion{Fe}{15} transitions (see, for example,
Austin et al. 1966; Widing \& Sandlin
1968; Behring, Cohen, \& Feldman 1972; Malinovsky \& Heroux 1973).  The most
detailed measurements of the {\em n} = 3--4 lines in \ion{Fe}{15} to
date were made by the XSST spectrograph during a rocket flight in 1982
(Acton et al. 1985).  However, even these observations have only been
subject to a brief analysis by Bhatia, Mason, \& Blancard (1997), as
part of their work to produce a large atomic dataset for this ion.
Finally, Campbell \& Brickhouse (2001) presented preliminary results
from an investigation of \ion{Fe}{15} and \ion{Fe}{16} lines in {\it Chandra}
X-ray spectra of the late-type coronally-active stars Capella
(G1~III + G8~III) and Procyon (F5~IV).

In this paper we use the most recent atomic physics calculations for
\ion{Fe}{15} to generate soft X-ray line ratios for a wide range of
electron temperatures and densities applicable to coronal plasmas.  We
compare these in detail with the XSST observations of Action et
al. (1985), to assess the usefulness of the \ion{Fe}{15} transitions
as plasma diagnostics. However, in addition we analyse a {\em Chandra}
spectrum of Capella constructed from multiple observations.  This
co-added X-ray spectrum has one of the highest signal-to-noise ratios ever
achieved for an astrophysical coronal plasma in the $\sim$25--175 \AA\
soft X-ray region, hence
allowing us to perform an investigation of the \ion{Fe}{15} {\em n}
= 3--4 lines in an astronomical source other than the Sun. 

The Capella spectrum is also made freely available, to encourage
further research on the soft X-ray spectral region. For example, 
in conjunction with the published
XSST solar line list, the Capella data should be useful for reliably
identifying new emission features.
In addition, 
there are potentially many useful diagnostic lines in the soft X-ray wavelength
range, especially for solar flare-type plasmas (see, for example, Brown et al. 1986;
Feldman et al. 1992), and hence further exploration and assessment of these is 
warranted.

\section{Theoretical Line Ratios} 

The model ion for \ion{Fe}{15}
consisted of the
energetically lowest 53 fine-structure levels belonging to
the 3{\em s}$^{2}$, 3{\em s}3{\em p}, 3{\em p}$^{2}$, 3{\em s}3{\em d}, 
3{\em p}3{\em d},
3{\em d}$^{2}$, 3{\em s}4{\em s}, 3{\em s}4{\em p}, 3{\em s}4{\em d}, 
3{\em p}4{\em s} and 3{\em s}4{\em f} configurations.
Energies of these levels were obtained from
Churilov et al. (1985), Churilov, Levashov, \& Wyart
(1989),
Litzen \& Redfors (1987), Sugar \& Corliss (1985) and
Aggarwal, Keenan, \& Msezane (2003).

Electron impact excitation rates for transitions among all
the levels
discussed above were taken from Aggarwal et al. (2003).
For Einstein A-coefficients, the calculations of
Deb, Aggarwal, \& Msezane (1999) were employed for
allowed and intercombination lines. 
Radiative rates for forbidden transitions have also been
taken from the work of Deb et al., although the data were
not included in the published paper for conciseness. However
we note that they
are available from one of the authors (K.Aggarwal@qub.ac.uk)
on request. Our A-value results for the forbidden
transitions are, in general, very similar to those calculated by others, 
such as Bhatia et al. (1997).  
Proton impact excitation is only important for
transitions among the 3{\em s}3{\em p} $^{3}$P levels of
\ion{Fe}{15}, and in 
the present analysis we have employed
the calculations of
Landman \& Brown (1979).

Using the atomic data discussed above in conjunction with 
a recently updated version of the statistical equilibrium
code of Dufton (1977), relative \ion{Fe}{15}
level populations and hence
emission line strengths were calculated for a wide range of
electron temperatures
(T$_{e}$) and densities (N$_{e}$).
Details of the procedures involved and approximations
made may be found in Dufton, and Dufton et al. (1978).

In Figures 1--7 we plot the theoretical
emission line intensity ratios R$_{1}$ through 
R$_{11}$, which are defined in Table 1. These ratios are shown 
as a function of T$_{e}$ and/or N$_{e}$ for a range of
temperatures (T$_{e}$ = 10$^{6}$--10$^{6.9}$ K) over which
\ion{Fe}{15} has a fractional abundance in ionization
equilibrium of N(\ion{Fe}{15})/(Fe) $>$ 10$^{-3}$
(Mazzotta et al.
1998), and for values of N$_{e}$ (= 10$^{8}$--10$^{13}$ cm$^{-3}$)
appropriate to solar and stellar coronal plasmas.
In some instances, the ratios are predicted to be insensitive to
electron density over the N$_{e}$ interval considered
(for example, R$_{1}$), in which case values are only
plotted as a function of electron temperature at a single
density. Others, such as R$_{2}$, are weakly dependent
on density, and hence are shown for a limited range of
N$_{e}$ values. However those ratios which vary significantly
with T$_{e}$ and N$_{e}$, such as R$_{4}$, are plotted over the
full range of plasma parameters. We note that the R$_{12}$
ratio is not shown in the figures, as it
has the same temperature and density dependence
as R$_{5}$, owing to common upper
levels, but with

R$_{12}$ = 2.95 $\times$ R$_{5}$.

Given errors in the adopted atomic data of typically
$\pm$10\%\ (see the references above),
we would expect the
theoretical ratios to be accurate to better than $\pm$20\%.

The ratios in Figures 1--7 are given relative to the 59.40 \AA\
transition, as this feature is the cleanest and 
most reliably detected
\ion{Fe}{15}
emission line in the soft X-ray spectral region. This has been
checked by a search of line lists, and also by generating
synthetic spectra with the latest version
(4.2) of the {\sc chianti} database (Dere et al. 1997;
Young et al. 2003), which confirm that
no blending species is present which has a line intensity
greater than 2\%\ that of the \ion{Fe}{15} 59.40 \AA\
feature. 
However we note that theoretical ratios 
involving any line pair are available electronically
from one of the authors (FPK\footnote{F.Keenan@qub.ac.uk})
on request.

An inspection of Figures 1--7 reveals that several of
the ratios are
sensitive to variations in the electron temperature
and/or density. For
example, R$_{7}$ varies by a factor of about 2.0
between T$_{e}$ = 10$^{6}$ and 10$^{6.9}$ K, while
being insensitive to the adopted
electron density. Similarly, R$_{11}$
changes by a factor of 3.4 over the (relatively
narrow) density interval of N$_{e}$ = 10$^{9}$--10$^{11}$~cm$^{-3}$
at T$_{e}$ = 10$^{6.3}$ K.
Hence the ratios should, in principle,
provide useful T$_{e}$-- and/or N$_{e}$--diagnostics for the 
\ion{Fe}{15} emitting
region of a plasma.

We note that the present theoretical line ratios are generally
not excessively different from other recent calculations.
For example,
at T$_{e}$ = 10$^{6.3}$ K and N$_{e}$ =
10$^{8}$ cm$^{-3}$, we calculate R$_{1}$ = 0.58 compared to
R$_{1}$ = 0.68 from {\sc chianti} and 
R$_{1}$ = 0.38 from Bhatia et al. (1997).
Similarly, at these plasma parameters,
we find R$_{7}$ = 3.6 compared to
R$_{7}$ = 3.5 ({\sc chianti}) and R$_{7}$ = 3.3 (Bhatia et al.).
However a notable exception is the R$_{2}$
ratio, for which we estimate 
R$_{2}$ = 0.17 at 
T$_{e}$ = 10$^{6.3}$ K and N$_{e}$ =
10$^{8}$ cm$^{-3}$,
and Bhatia et al. calculate R$_{2}$ = 0.11. By contrast, {\sc chianti}
indicates that R$_{2}$ = 0.019, a factor of 6--9
lower than the other results.
This discrepancy arises due to the adopted
A-value for the 
3{\em s}$^{2}$ $^{1}$S--3{\em s}4{\em p} $^{3}$P$_{1}$
transition. In our calculations, we use A = 8.1$\times$10$^{10}$ s$^{-1}$
from Deb et al. (1999), and Bhatia et al. employ their
own value of A = 8.4$\times$10$^{10}$ s$^{-1}$.
However the {\sc chianti}
database uses A = 1.5$\times$10$^{10}$ s$^{-1}$
from Griffin et al. (1999), which is significantly
smaller than the other two calculations.
Furthermore, Deb et al. undertook 
two independent calculations of radiative rates for 
\ion{Fe}{15}, using the CIV3 atomic structure code of Hibbert (1975)
and the GRASP code of Dyall et al. (1989). Their A-values for 
the 3{\em s}$^{2}$ $^{1}$S--3{\em s}4{\em p} $^{3}$P$_{1}$
transition from the two codes 
differ by only 11\%, and as a consequence 
they assess that their 
calculations should be accurate to better than 20\%.
Given this, and the excellent agreement with the result of 
Bhatia et al., who also performed an independent calculation
with the SUPERSTRUCTURE code of Eissner et al. (1972), we 
therefore believe that the Griffin et al. 
A-value is in error and should be re-evaluated.
  
\section{Observational Data}

We compare our \ion{Fe}{15} line ratio calculations with two datasets,
namely that for a solar flare obtained with a rocket-borne
spectrograph, and a composite spectrum for Capella observed with the
{\em Chandra} satellite. These datasets are discussed separately below.

\subsection{Solar flare observations}

The solar spectrum analysed in the present paper is that
of an M-class flare, recorded on Kodak 101--07
emulsion by an X-ray spectrograph (XSST) during a rocket flight
on 1982 July 13. These observations spanned the wavelength range
11--97 \AA, at a spectral resolution of 0.02 \AA, and covered a solid
angle of approximately 625 arcsec$^{2}$, centered on or near the brightest
source of X-rays.
The XSST made two exposures, one of 54 s beginning at
16:33:50 UT and the other of 145 s commencing at 16:35:35 UT.
Unfortunately, a failure of the mechanism to move the film more 
than one mm between the two exposures 
resulted in their being nearly superimposed on the same plate.  
However, a careful analysis
 allowed the 
two spectra to be separated, so that analysis of the data
could be performed, and the observations presented here
are for the 145 s exposure.
Further details of the XSST instrument may be found in
Brown et al. (1979) and Bruner et al. (1980),
while the observations are discussed in Acton et al. (1985).

It would clearly have been preferable to model the XSST 
observations in detail using profile fitting methods, as for the {\em Chandra}
data (see \S\ 3.2). This was unfortunately  
not possible as the original
XSST spectra are no longer available, following the retirement of
several key staff involved in the mission.
Consequently, only the published line list and intensities for XSST are now accessible.
We have searched for \ion{Fe}{15} emission lines in the
XSST spectrum, using the identifications of Acton et al. (1985)
and also the NIST database\footnote{http://physics.nist.gov/PhysRefData/}
and other line lists, such as the Atomic Line List of 
Peter van Hoof\footnote{http://star.pst.qub.ac.uk/$\sim$pvh/}.
In Table 1 we list the \ion{Fe}{15}
transitions found in the spectrum, along with their measured wavelengths.

The intensity of the 59.40 \AA\ line of \ion{Fe}{15} measured by Acton
et al. (1985) in the XSST spectrum is given in Table 2; observed
intensities of the other \ion{Fe}{15} transitions may be inferred from
this using the line ratios given in the table (see \S\ 2).  
Acton et al. (1985) note that a strong second-order 
spectrum falling between 25--50 \AA\ is evident in the XSST observations, and that the
effects of scattered light and double exposure 
made the reduction of the data difficult. However although difficult, 
Acton et al. did succeed in producing a well-calibrated spectrum and 
hence reliable measurements of emission line intensities.
Evidence for this comes from, for example, our previous
analysis of \ion{Ni}{18} emission lines in the XSST spectrum (Keenan et al. 1999).
We found that, for 6 out of 7 line ratios involving \ion{Ni}{18}
transitions in the 41--53 \AA\ wavelength region, agreement between theory and observation
is excellent, with differences that average only 11\%. 
For the sole \ion{Ni}{18}
line ratio which shows a large discrepancy between theory and observation, 
this is explained by a known blend with a \ion{Fe}{19}
transition. The \ion{Fe}{15} lines in the XSST spectrum should in fact
be more reliably measured than those for \ion{Ni}{18}, as the former lie outside
the 25--50 \AA\ range affected by the second-order spectrum. 
Brown et
al. (1986) and Keenan et al. note that the relative intensities
of lines in the XSST spectrum similar in strength to the \ion{Fe}{15}
transitions discussed here should be accurate to about $\pm$20\%, and
hence line ratios to $\pm$30\%. The observed \ion{Fe}{15} line ratios
in Table 2 have therefore been assigned a uniform $\pm$30\%\
uncertainty.

\subsection{Capella Spectrum}

\subsubsection{Observations and Reduction}

Capella is the mainstay target for calibrating and monitoring the {\em
Chandra} transmission grating spectrometer dispersion relation.  As
such, it is currently observed approximately every 12 months using the
LETG+HRC-S combination.  The Capella observations analysed here were
obtained by the {\em Chandra} Low Energy Transmission Grating
Spectrograph (LETGS), employing the High Resolution Camera
spectroscopic detector (HRC-S) in its standard configuration, yielding
spectra with a resolution of approximately 0.06~\AA.  (See Weisskopf
et al. 2003 for a recent overview of the {\em Chandra} instrumentation
and its in-flight performance).  We have taken advantage of the
multiple observations of Capella accrued to date to construct a
composite spectrum with the highest signal-to-noise ratio possible.
The different observations utilised, together with their dates of
acquisition and exposure times, are listed in Table 3.  Observations
from 2002 were obtained at significantly different off-axis angles,
and were not included in this analysis.

Data were uniformly processed and spectra extracted using the {\em
Chandra} Interactive Analysis of Observations (CIAO) software Version
3.1.  Photon event lists were also filtered in detector pulse height
in order to remove particle-induced background events\footnote{See
http://cxc.harvard.edu/cal/Letg/Hrc\_bg/}.  The final extracted
co-added spectrum is shown in Figure 8, and we note that this is
freely available from one of the authors
(JJD\footnote{jdrake@cfa.harvard.edu}) electronically on request. In
Figure 9 we plot the 50--75 \AA\ portion of the spectrum, as this
contains all of the \ion{Fe}{15} soft X-ray lines identified in
Capella. Unfortunately, the spectrum is too noisy at longer
wavelengths for other (weak) \ion{Fe}{15} features to be detected.

\subsubsection{Analysis}

The Capella spectrum was analysed using the PINTofALE\footnote{Freely
available from http://hea-www.harvard.edu/PINTofALE/}
IDL\footnote{Interactive Data Language, Research Systems Inc.}
software suite (Kashyap \& Drake 2000).  Line fluxes were measured by
fitting ``modified Lorentzian'', or Moffat,  functions of the form
$F(\lambda)=a/(1+\frac{\lambda-\lambda_0}{\Gamma})^\beta$, where $a$
is the amplitude and $\Gamma$ a characteristic line width.  For a
value of $\beta=2.4$, it has been found that $F(\lambda)$ reproduces
the line response function of the LETG+HRC-S instrument to the
photometric accuracy of lines with a few thousand counts or less
(Drake 2004).  Uncertainties were estimated using a Monte Carlo
sampling of the free parameters in the line fits.

Free parameters are, in general, the line width, position and
amplitude, although for some of the fitting of poorer quality spectral
features we constrained the line width to the value 0.065 \AA\ found
for nearby clean lines.  Continua were estimated locally by eye for
each fit.  Line positions were allowed to vary from their reference
positions by $\leq$0.05 \AA, this being dictated by the imaging
characteristics of the HRC-S detector.  As discussed by Chung et
al. (2004), the HRC-S exhibits small-scale imaging non-linearities
that vary over the detector, and which can displace a spectral line
from its true position by up to 0.05 \AA .  While this effect can be
calibrated to some extent using bright, well-known spectral lines, the
wavelength range of interest to this \ion{Fe}{15} study is devoid of
such transitions, and the characteristics of the imaging
non-linearities have not been well defined in the relevant regions of
the detector.  For lines closely spaced in wavelength, the relative
separations of the lines were kept fixed to their reference values,
while the position of the group was allowed to vary.  In the case of
the group of lines in the 69.4--70.2 \AA\ range (Figure 10), the
wavelengths of the 69.93, 69.98 and 70.05 \AA\ transitions were
decoupled from the stronger 69.65 \AA\ feature. We found the best-fit
location of the former group to be displaced by 0.04 \AA\ from the
reference position, while the latter line lies very close to its
expected location.

The relatively unexplored and complex nature of the crowded soft X-ray
spectrum in the 50--75 \AA\ wavelength range poses particular
challenges for estimating line fluxes.  In addition to the lines from
the $n=2$ shells of abundant elements such as Mg, Si, S and Ar, and
the $n=3$ shell of Fe, lines from shorter wavelengths arising from
higher spectral orders are also present in this range.  Unlike the
{\it Chandra} ACIS CCD detector, the HRC-S microchannel plate detector
possesses no energy resolution of its own and overlapping spectral orders
cannot be separated.  Prior to performing line fits we therefore
looked for the presence of significant blends from known strong lines
in higher orders.  

The LETGS is designed with a grating bar-to-space ratio of 1:1, which
results in some suppression of even orders.  The most important
order for line blends in the 50--75 \AA\ range is the 3rd, which has
an efficiency of about 10\%\ that of 1st order.\footnote{see
http://cxc.harvard.edu/cal/Letg/HO2004/} Two of the \ion{Fe}{15} lines
in our list coincided reasonably closely with noteable 3rd order
lines: $\lambda 66.25$, which lies slightly less than one line width
away from 3rd order \ion{O}{7} $\lambda 22.098$; and $\lambda 52.91$,
which lies a line width away from 3rd order \ion{Fe}{18}~$\lambda
17.62$.  The \ion{O}{7} forbidden line is quite strong in the spectrum
of Capella, and \ion{Fe}{15} $\lambda 66.25$ was therefore discarded.
However \ion{Fe}{18} $\lambda 17.62$ is a much weaker line, of which
there is no obvious sign in 3rd order.  It arises between the levels
$2s^{2}2p^{4}3p$\space$^{2}P_{3/2}$ and $2s2p^{6}$\space$^{2}S_{1/2}$,
and occurs because of configuration mixing (Drake et al.\ 1999).

Spectral orders 4 and 5 have similar efficiencies of 2--3~\%\ that 
for 1st
order.  The \ion{O}{8} + \ion{Fe}{18} $\lambda 16.00$ blend coincides
in 4th order with \ion{Fe}{15}~$\lambda 63.97$, and was included in the
spectral fit with the relative wavelengths of the two components kept
fixed.  While there are no other strong lines blended with our
\ion{Fe}{15} features in 4th and 5th orders, there is a slew of weaker
lines from $\Delta n >0$ transitions in the {\em n} = 
2 shells of Fe and Ni
that fall in the 50--75~\AA\ range.  These transitions give rise to an
additional pseudocontinuum, that we treated empirically through
the estimation of a local continuum for each line or line group.
Such empirical local continua were also guided by the broad-band ``by
eye'' continuum, that can be seen in Figure 8.

In the fitting of some lines, neighbouring features blend in with the
line wings.  In most cases, these blending features are not clean
lines that could be unambiguously identified with particular
transitions of accurately known wavelengths, but are often complex
blends with some unidentified contributions.  To account for these
unidentified transitions in the measured fluxes of our lines of
interest, we added line components in an ad hoc fashion until the
profiles of blended features could be empirically matched.  Examples
are shown in Figure 10, while the observed \ion{Fe}{15} line ratios
are listed in Table 4, along with their associated errors.  In Table
5 we list the wavelengths of the additional ad hoc line components.

As with
the XSST observations (\S\ 3.1), the measured intensity of the 59.40
\AA\ line is given in Table 4, and those of the other \ion{Fe}{15}
transitions may be inferred from this using the line ratio values.  It
should be kept in mind in the interpretation of the measured fluxes
that they are also prone to uncertainty caused by hidden blends of
unidentified lines.  For such cases, there is an expectation that the
measured fluxes might be systematically too high.

\section{Results and Discussion}

In Tables 2 and 4 we list the observed \ion{Fe}{15} emission line
ratios from the XSST solar flare spectrum and the {\em Chandra}
Capella observations, respectively.  Also shown in the tables are the
theoretical results from Figures 1--7 at the temperature of maximum
fractional abundance in ionization equilibrium for \ion{Fe}{15},
T$_{e}$ = 10$^{6.3}$ K (Mazzotta et al. 1998), and at electron
densities derived for the solar and Capella observations from emission
line ratios in \ion{O}{7} (Brown et al. 1986; Phillips et
al. 2001). The temperature-sensitive $G$-ratio for \ion{O}{7} in the
XSST and Capella data indicates that T$_{e}$ = 10$^{6.3}$ K in both
features, similar to that of maximum fractional abundance for
\ion{Fe}{15}.  In the case of Capella, whose coronal emission measure
distribution rises quite steeply to a peak at T$_{e}$ =
6$\times$10$^6$~K (see, for example, 
Brickhouse et al. 2000), it is possible that this $G$-ratio
slightly underestimates the mean temperature of line formation, as was
found empirically by Testa, Drake \& Peres (2004) for a sample of
active stars.  However, the ion populations of both \ion{O}{7} and
\ion{Fe}{15} peak at very similar temperatures, and hence the
\ion{O}{7} density should still accurately reflect that of the
\ion{Fe}{15} emitting region of the plasma.  We have also verified
that the slightly higher temperatures of line formation that would
result from the relatively steep Capella emission measure distribution
do not significantly change the predicted \ion{Fe}{15} line ratios
from those listed in Table~4.  The error bars on the theoretical
results are based on the estimated $\pm$20\%\ accuracy of the line
ratio calculations (see \S\ 2).

An inspection of Table 2 reveals very good agreement between the
XSST observations and the theoretical predictions
for the ratios  R$_{5}$ and R$_{7}$.
This indicates that the 63.97 and 69.65 \AA\ lines are 
well detected in the XSST spectrum,
and must be relatively free from blends. In particular, we confirm 
the identification by Bhatia et al. of the 63.97 \AA\ feature 
as the 3{\em p}$^{2}$ $^{1}$D--3{\em s}4{\em f} $^{1}$F 
transition of \ion{Fe}{15}, and not the 
\ion{Al}{8} 2{\em s}$^{2}$2{\em p}$^{2}$ $^{3}$P$_{1,2}$--2{\em 
s}2{\em p}$^{2}$3{\em p} $^{3}$D$_{2,3}$ lines 
as originally classified by Acton et al. (1985). 
These authors also identify the 69.65 \AA\ 
line as being partially due to \ion{Si}{8} and \ion{Fe}{14}
transitions,
with an unknown feature responsible for most of the line
flux. However our result for R$_{7}$ clearly shows that the
line is primarily the \ion{Fe}{15} 3{\em s}3{\em p} $^{1}$P--3{\em s}4{\em s} 
$^{1}$S transition, as suggested by Bhatia et al.
We have confirmed this by generating a synthetic spectrum 
using {\sc chianti}, which indicates that the 
\ion{Si}{8} 2{\em s}$^{2}$2{\em p}$^{3}$ $^{4}$S--2{\em s}$^{2}$2{\em p}$^{2}$3{\em s}
$^{4}$P$_{5/2}$ line only contributes
about 15\%\ to the total 69.65 \AA\ flux, and that
no \ion{Fe}{14} transitions are present.  

Similarly, the observed R$_{1}$, R$_{9}$, R$_{10}$, R$_{11}$
and R$_{12}$
ratios in Table 2 
are only slightly larger than the theoretical values,
indicating that the 52.91, 69.98, 70.05, 73.47 and 82.76 \AA\
lines must be mostly due to \ion{Fe}{15}. Indeed, within the
observational and theoretical errors in the line ratios,
all the observed fluxes may arise from this ion.
This is supported by the {\sc chianti}
synthetic spectrum, which predicts no significant blending
species for the 52.91, 69.98 and 70.05 \AA\ features, and contributions
of only 
about 10\%\ from \ion{Ne}{8} 1{\em s}$^{2}$2{\em p} $^{2}$P$_{1/2}$--1{\em 
s}$^{2}$4{\em d} $^{2}$D$_{3/2}$ and 20\%\ from \ion{Ne}{9}
1{\em s}2{\em p} $^{1}$P--1{\em s}3{\em s} $^{1}$S to
the 73.47 and 82.76 \AA\ lines, respectively.
Our classification of the 82.76 \AA\ feature as the 
\ion{Fe}{15} 3{\em s}3{\em d} $^{3}$D$_{3}$--3{\em 
s}4{\em p} $^{3}$P$_{2}$ line 
is, to our knowledge, the first time this transition has been
identified in an astrophysical source.

The good agreement between theory and observation for the above 
line ratios indicates that they should provide useful 
diagnostics for the \ion{Fe}{15}
emitting region of a plasma. 
In particular, R$_{7}$, R$_{9}$, R$_{10}$ and R$_{11}$
are all temperature and/or density sensitive, and hence with careful
use should allow these parameters to be derived. For example, R$_{7}$
could be employed to evaluate T$_{e}$, as it does not vary
with N$_{e}$ (Figure 1), and the latter could then be determined
from R$_{10}$ or R$_{11}$ (Figures 6 and 7).

The observed values of R$_{2}$, R$_{3}$, R$_{4}$, R$_{6}$ and R$_{8}$
in the XSST spectrum are all significantly larger
than theory, implying that the 53.11, 55.78, 56.17, 66.25 and 69.93 \AA\
lines of \ion{Fe}{15}
are blended to varying degrees.
For the 53.11 \AA\ feature, the blending 
line contributes about 50\%\ of the measured flux, and the 
{\sc chianti}
synthetic spectrum indicates that this is most likely the
\ion{S}{9}  2{\em s}$^{2}$2{\em p}$^{4}$ $^{3}$P$_{1}$--2{\em 
p}$^{3}$3{\em s} $^{3}$P$_{2}$ transition. This is predicted to 
have an intensity of approximately 10\%\ that of the \ion{Fe}{16}
3{\em p} $^{2}$P$_{1/2}$--4{\em 
d} $^{2}$D$_{3/2}$ line at 54.13 \AA, i.e. about 16 photons cm$^{-2}$ s$^{-1}$
arcsec$^{-2}$.
The measured flux of the 53.11 \AA\ feature is 31 photons 
cm$^{-2}$ s$^{-1}$
arcsec$^{-2}$, so \ion{S}{9} should 
contribute 50\%, in agreement with observation. We note that
Acton et al. (1985) did not classify the 
53.11 \AA\ feature, the \ion{Fe}{15} identification 
being suggested by Bhatia et al. (1997).
However we can now confirm that the line is 
the 3{\em s}$^{2}$ $^{1}$S--3{\em s}4{\em p} $^{3}$P$_{1}$
transition of \ion{Fe}{15}, 
although it is blended. 

For the 55.78 \AA\ line, {\sc chianti}
indicates that the \ion{Si}{10} 
2{\em s}2{\em p}$^{2}$ $^{2}$S--2{\em s}2{\em p}3{\em s} $^{2}$P$_{1/2,3/2}$
and \ion{Si}{9} 2{\em s}$^{2}$2{\em p}$^{2}$ $^{3}$P$_{1}$--2{\em s}$^{2}$2{\em 
p}3{\em d} $^{1}$D transitions should together have an intensity
about 70\%\
that of the \ion{Fe}{15} line. Our observation of R$_{3}$
implies that the blending species
may be stronger, and contribute about 150\%\ of the \ion{Fe}{15} flux. 
However when the uncertainties 
in the experimental and theoretical 
R$_{3}$ ratios are taken into account, the data are compatible with a 70\%\
contribution from \ion{Si}{10} and \ion{Si}{9}.  
Similarly, a comparison of theory and observation for R$_{6}$
indicates that \ion{Fe}{15} only contributes about 10\%\ 
to the intensity of the 66.25 \AA\ feature. This is in agreement with
{\sc chianti}, which predicts that 90\%\ of the observed flux is due to 
the \ion{Fe}{16} 3{\em d} $^{2}$D$_{3/2}$--4{\em f} $^{2}$F$_{5/2}$
line. 

By contrast, it is not clear from {\sc chianti}
what the blending species are for the 56.17 and 69.93 \AA\ lines, with
no transitions predicted to have intensities more than 10\%\ those of the
relevant \ion{Fe}{15} features.
It is however possible that the blending may be due to strong lines
observed by XSST in second- or third-order, which are appearing in 
first-order at these wavelengths.
For example, in the case of the 56.17 \AA\ line, {\sc chianti}
indicates no strong transitions in second-order (i.e. around
28.09 \AA), but the intense \ion{Ca}{18} 1{\em s}$^{2}$2{\em s} 
$^{2}$S--1{\em s}$^{2}$3{\em p} $^{2}$P$_{1/2}$ transition is
predicted in third-order (18.73 \AA). Similarly, for the 69.93 \AA\
line there is a predicted third-order feature at 23.31 \AA, namely
the \ion{Ca}{16} 2{\em s}2{\em p}$^{2}$ 
$^{2}$P$_{3/2}$--2{\em s}2{\em p}3{\em d} $^{2}$D$_{5/2}$
transition. There are unfortunately no \ion{Ca}{18} lines
in the XSST spectrum for comparison purposes, but there is
a \ion{Ca}{16} transition at 21.45 \AA\ (2{\em s}$^{2}$2{\em p} 
$^{2}$P$_{1/2}$--2{\em s}$^{2}$3{\em d} $^{2}$D$_{3/2}$),
with an intensity I = 37 photons 
cm$^{-2}$ s$^{-1}$
arcsec$^{-2}$. However the
predicted {\sc chianti}
intensity ratio is I(23.31 \AA)/I(21.45 \AA) $<$ 10$^{-3}$, indicating that
the \ion{Ca}{16} 23.31 \AA\ transition cannot be responsible for 
the blend in the
69.93 \AA\ line. Clearly, the identification of the blends for the 56.17 and
69.93 \AA\ features will require
further work.

In the case of the {\em Chandra} observations of Capella, agreement
between theory and experiment in Table 4 is very good for the majority
of the line ratios. Where there are discrepancies, such as for
R$_{8}$, these tend to mirror those found for the XSST spectrum, with
the observed value being somewhat larger than theory, indicative of
some blending.  However, in most instances the observed line ratios
actually show smaller discrepancies with theory than do the XSST
results. To a certain extent, such agreement may be judged to be
unsurprising, as in our analysis of the Capella observations we added
line components in an ad hoc fashion until the profiles of blending
features could be empirically matched (see \S\ 3.2). Nevertheless, our
results indicate that profile fitting of the {\em Chandra} data,
rather than simple measurements of emission line intensities, does
allow the reliable detection of \ion{Fe}{15} soft X-ray features in
the 50--75 \AA\ wavelength range. Indeed, our work represents (to our
knowledge) the first time that the \ion{Fe}{15} {\em n} = 3--4 lines
have been identified in an astrophysical source other than the
Sun. Hopefully, our results will encourage the future use of these
lines as plasma diagnostics for this spectral region.

In summary, we have performed a detailed analysis of \ion{Fe}{15} soft X-ray 
lines in the $\sim$52--83 \AA\ wavelength region, observed in the solar spectrum 
by the XSST spectrograph, and confirmed identifications
for several emission features and also detected a new transition.  
Additionally, the \ion{Fe}{15} lines have been measured 
in {\em Chandra} observations of Capella, and appear to provide potentially
useful
plasma diagnostics. We therefore hope that our work will stimulate 
renewed interest in this relatively unexplored
spectral region, in particular through our
provision of the co-added {\em Chandra} spectrum for Capella, which are
probably the highest 
signal-to-noise data available. Many dozens of lines in the solar 
soft X-ray spectrum are without
identifications, and the Capella spectrum (in combination with the XSST solar 
line list) 
should provide an aid to the identification of soft X-ray transitions and the
investigation of their usefulness as plasma diagnostics. 
This will be important for future space missions, and in particular 
the EUV Variability Experiment (EVE), which is part of the
{\em Solar Dynamics Observatory} (SDO), due for launch in 2008
\footnote{See http://lasp.colorado.edu/eve/}.
One of the EVE instruments is the Multiple EUV Grating Spectrograph (MEGS), which
will observe the 50--1050 \AA\ region at a spectral resolution of 1 \AA.
Clearly, the lines in this wavelength range will need to be identified and
understood in advance of MEGS observations, as this instrument will not be
able to resolve them individually.

\acknowledgements

K.M.A., R.S.I.R. and D.S.B. acknowledge financial support from the 
EPSRC and PPARC 
Research Councils of the
United Kingdom. 
F.P.K. is grateful to AWE Aldermaston for the award of a William Penney
Fellowship. The authors thank Peter van Hoof for the use of his
Atomic Line List, and Jeffrey Linsky for very useful comments on an earlier
version of the paper.
{\sc chianti} is a collaborative project 
involving the Naval Research Laboratory (USA), Rutherford
Appleton Laboratory (UK), and the Universities of Florence 
(Italy) and Cambridge (UK).

\clearpage


\begin{figure}
\includegraphics[scale=0.4,angle=90]{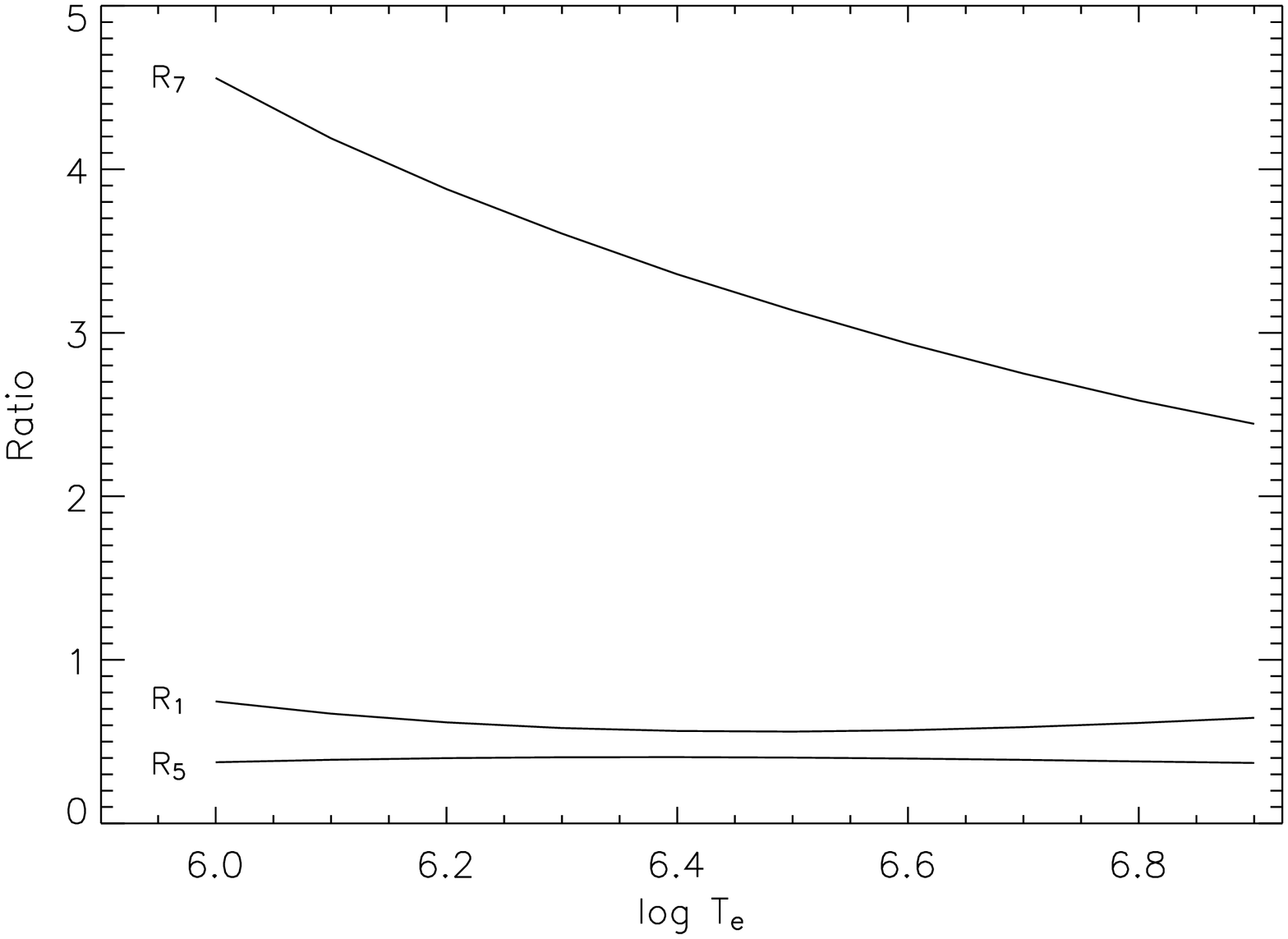}
\caption{
The theoretical \ion{Fe}{15}
emission line intensity ratios
R$_{1}$ = I(52.91 \AA)/I(59.40 \AA), R$_{5}$ = I(63.97 \AA)/I(59.40 \AA) 
and R$_{7}$ = I(69.65 \AA)/I(59.40 \AA),
where I is in photon units,
plotted as a function of logarithmic electron temperature
(T$_{e}$ in K) at an electron density of
N$_{e}$ = 10$^{10}$ cm$^{-3}$. However we note that
the line ratios
are insensitive to the adopted electron density for N$_{e}$
= 10$^{8}$--10$^{13}$ cm$^{-3}$.  
}
\end{figure}

\begin{figure}
\includegraphics[scale=0.4,angle=90]{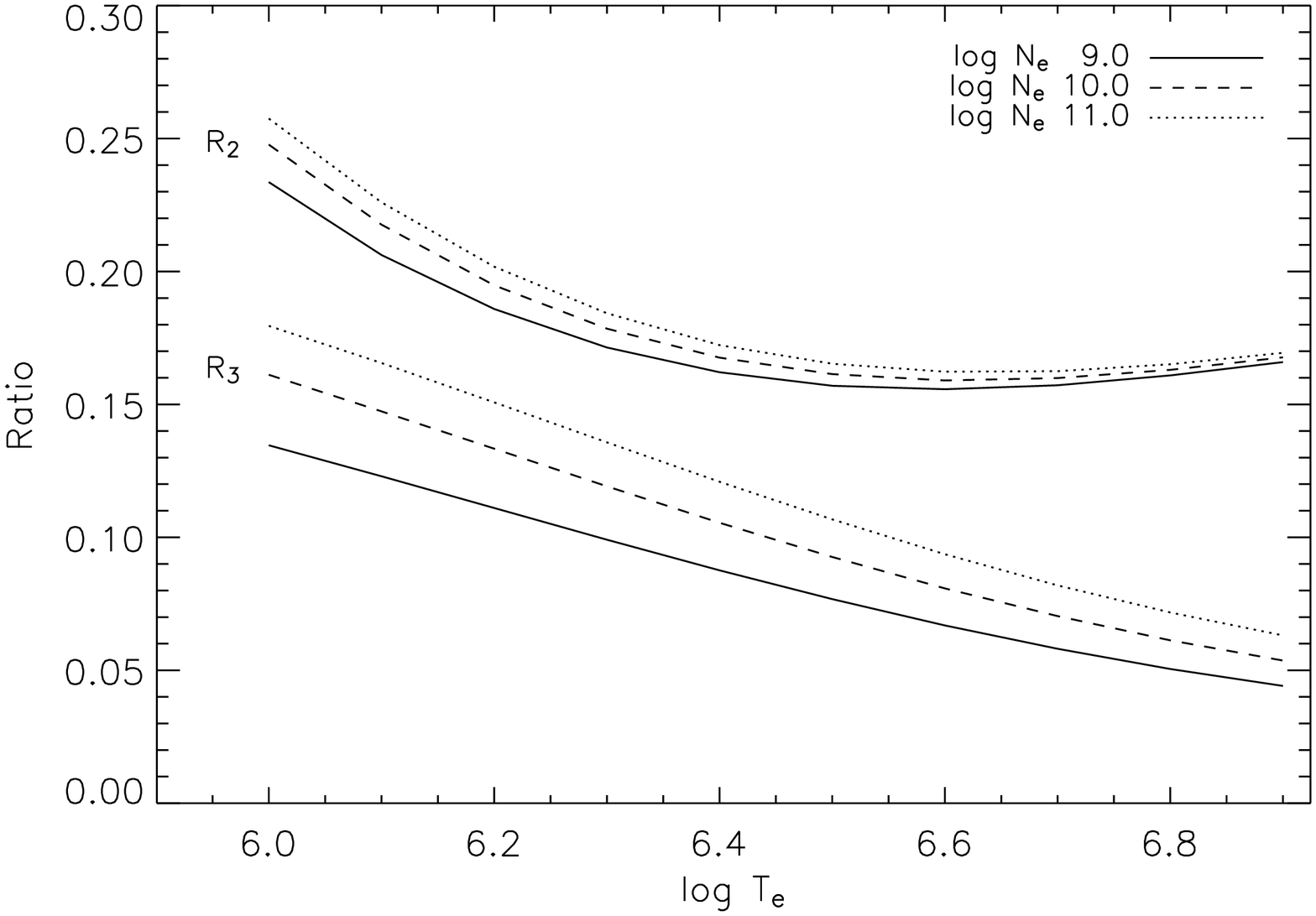}
\caption{
The theoretical \ion{Fe}{15}
emission line intensity ratios
R$_{2}$ = I(53.11 \AA)/I(59.40 \AA) 
and R$_{3}$ = I(55.78 \AA)/I(59.40 \AA),
where I is in photon units,
plotted as a function of logarithmic electron temperature
(T$_{e}$ in K) at logarithmic electron densities
of log N$_{e}$ = 9.0, 10.0 and 11.0 (N$_{e}$ in cm$^{-3}$).  
}
\end{figure}

\begin{figure}
\includegraphics[scale=0.4,angle=90]{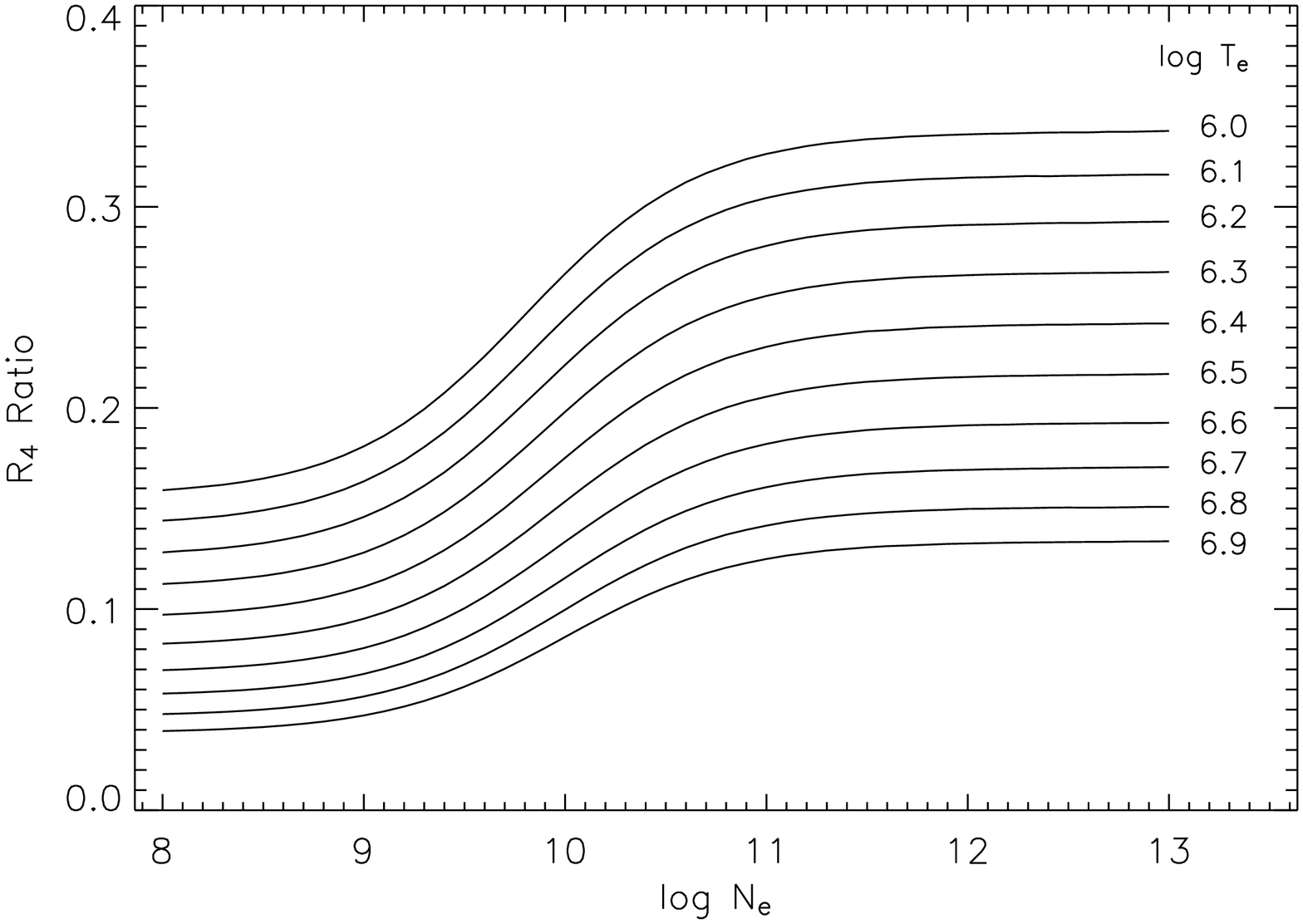}
\caption{
The theoretical \ion{Fe}{15}
emission line intensity ratio
R$_{4}$ = I(56.17 \AA)/I(59.40 \AA),
where I is in photon units,
plotted as a function of logarithmic electron density
(N$_{e}$ in cm$^{-3}$) for several values of logarithmic
electron temperature in the range log T$_{e}$ = 6.0--6.9 (T$_{e}$
in K).  
}
\end{figure}

\begin{figure}
\includegraphics[scale=0.4,angle=90]{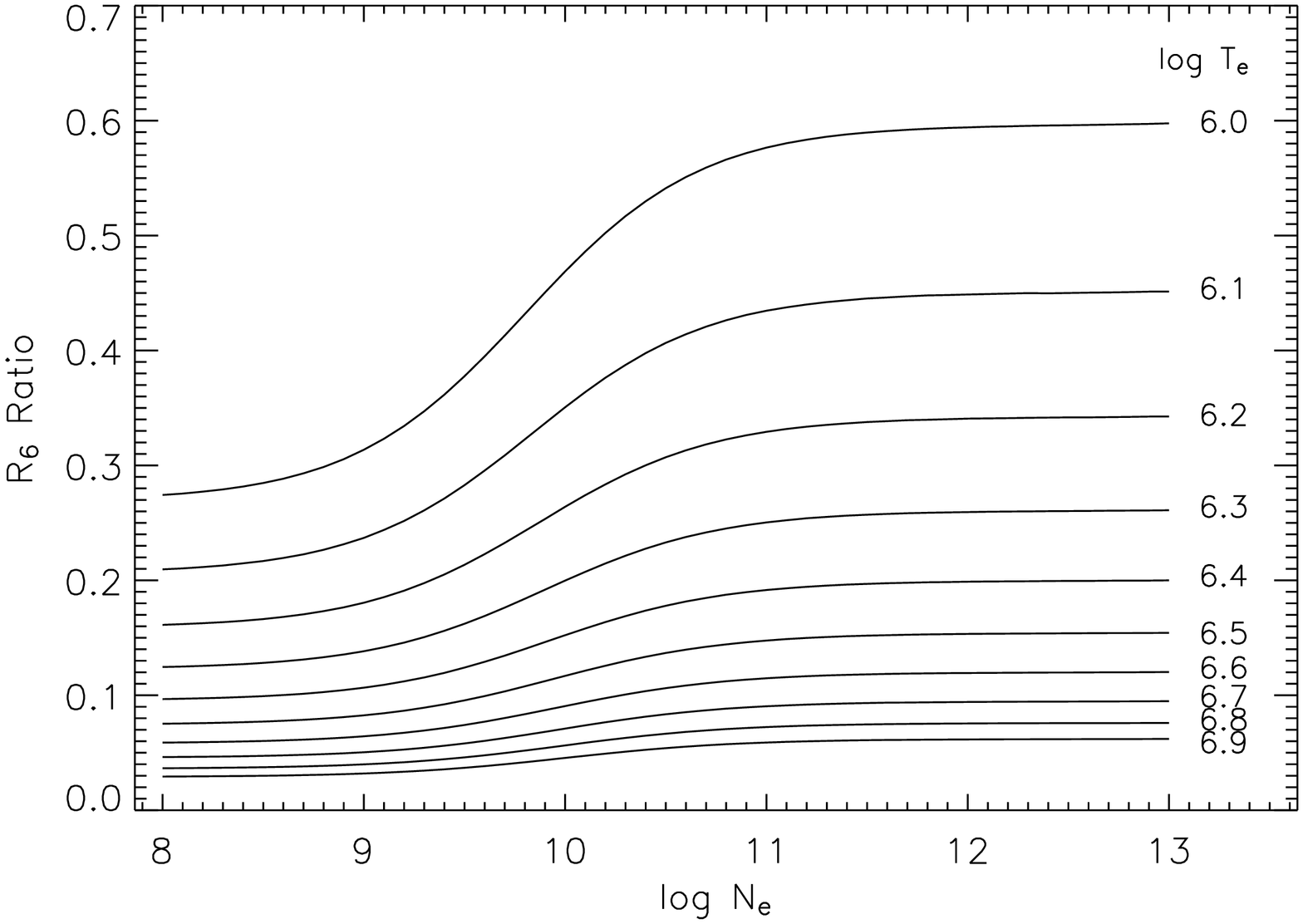}
\caption{
The theoretical \ion{Fe}{15}
emission line intensity ratio
R$_{6}$ = I(66.25 \AA)/I(59.40 \AA),
where I is in photon units,
plotted as a function of logarithmic electron density
(N$_{e}$ in cm$^{-3}$) for several values of logarithmic
electron temperature in the range log T$_{e}$ = 6.0--6.9 (T$_{e}$
in K).  
}
\end{figure}

\begin{figure}
\includegraphics[scale=0.4,angle=90]{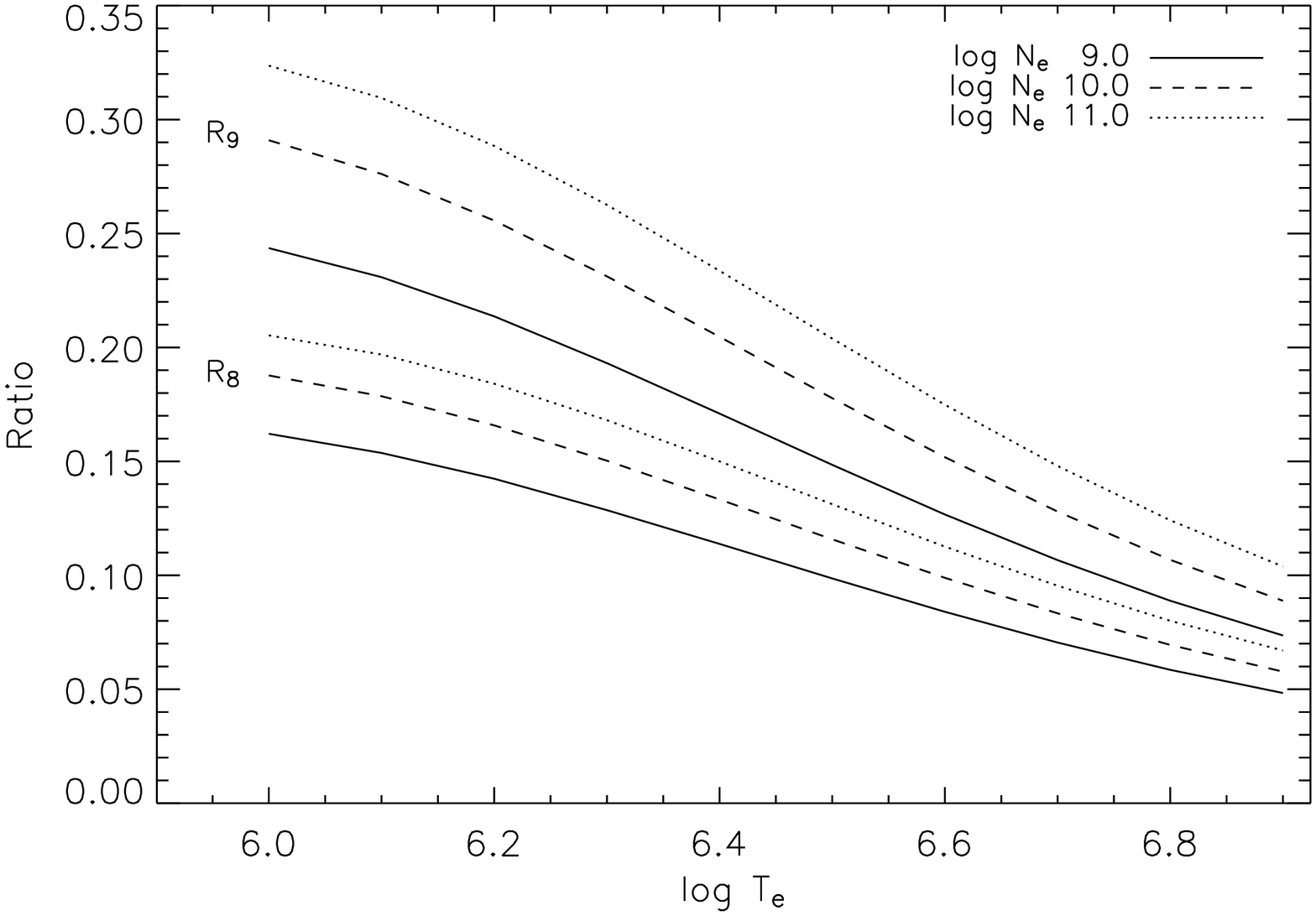}
\caption{
The theoretical \ion{Fe}{15}
emission line intensity ratios
R$_{8}$ = I(69.93 \AA)/I(59.40 \AA) 
and R$_{9}$ = I(69.98 \AA)/I(59.40 \AA),
where I is in photon units,
plotted as a function of logarithmic electron temperature
(T$_{e}$ in K) at logarithmic electron densities
of log N$_{e}$ = 9.0, 10.0 and 11.0 (N$_{e}$ in cm$^{-3}$).  
}
\end{figure}

\begin{figure}
\includegraphics[scale=0.4,angle=90]{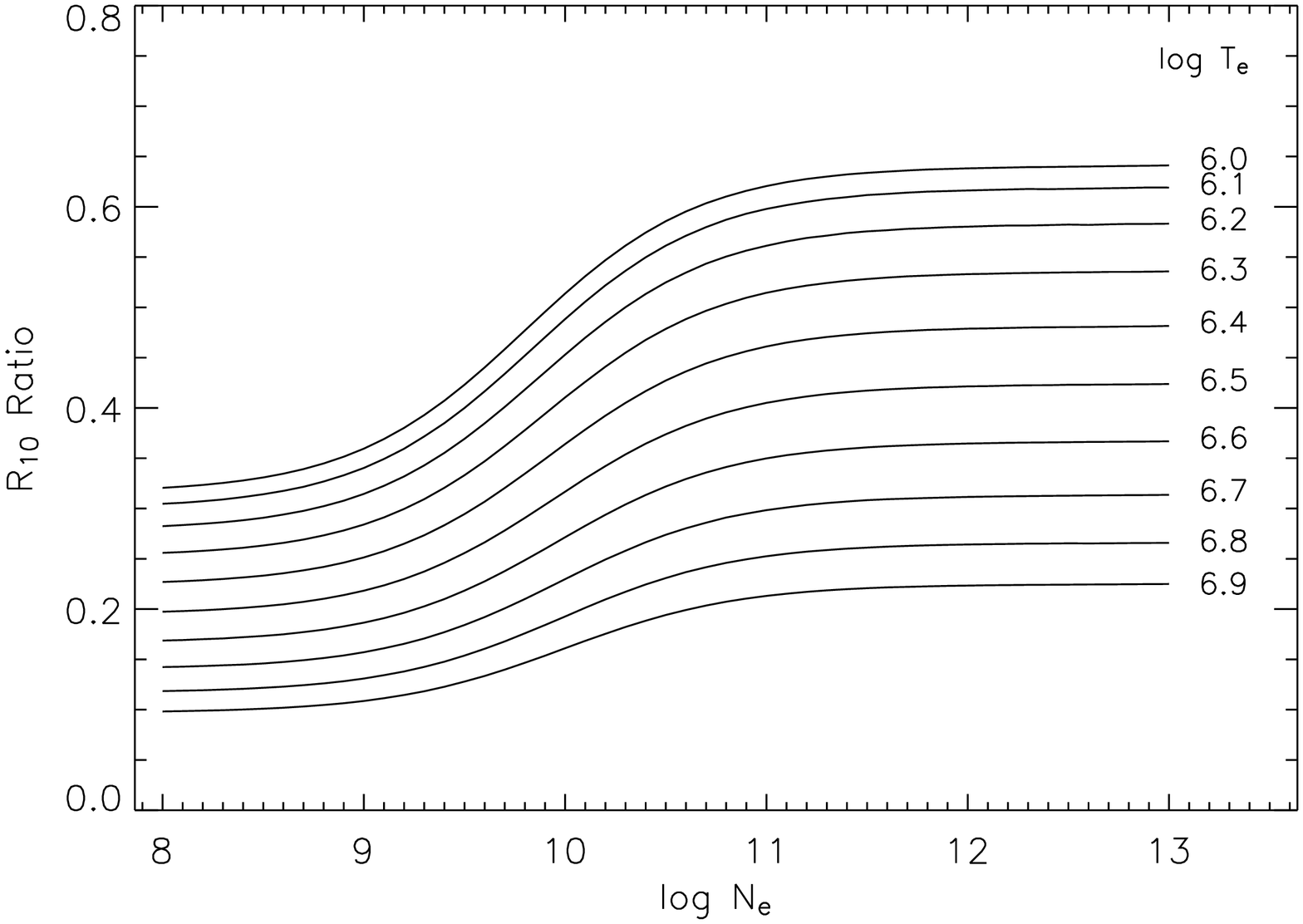}
\caption{
The theoretical \ion{Fe}{15}
emission line intensity ratio
R$_{10}$ = I(70.05 \AA)/I(59.40 \AA),
where I is in photon units,
plotted as a function of logarithmic electron density
(N$_{e}$ in cm$^{-3}$) for several values of logarithmic
electron temperature in the range log T$_{e}$ = 6.0--6.9 (T$_{e}$
in K).  
}
\end{figure}

\begin{figure}
\includegraphics[scale=0.4,angle=90]{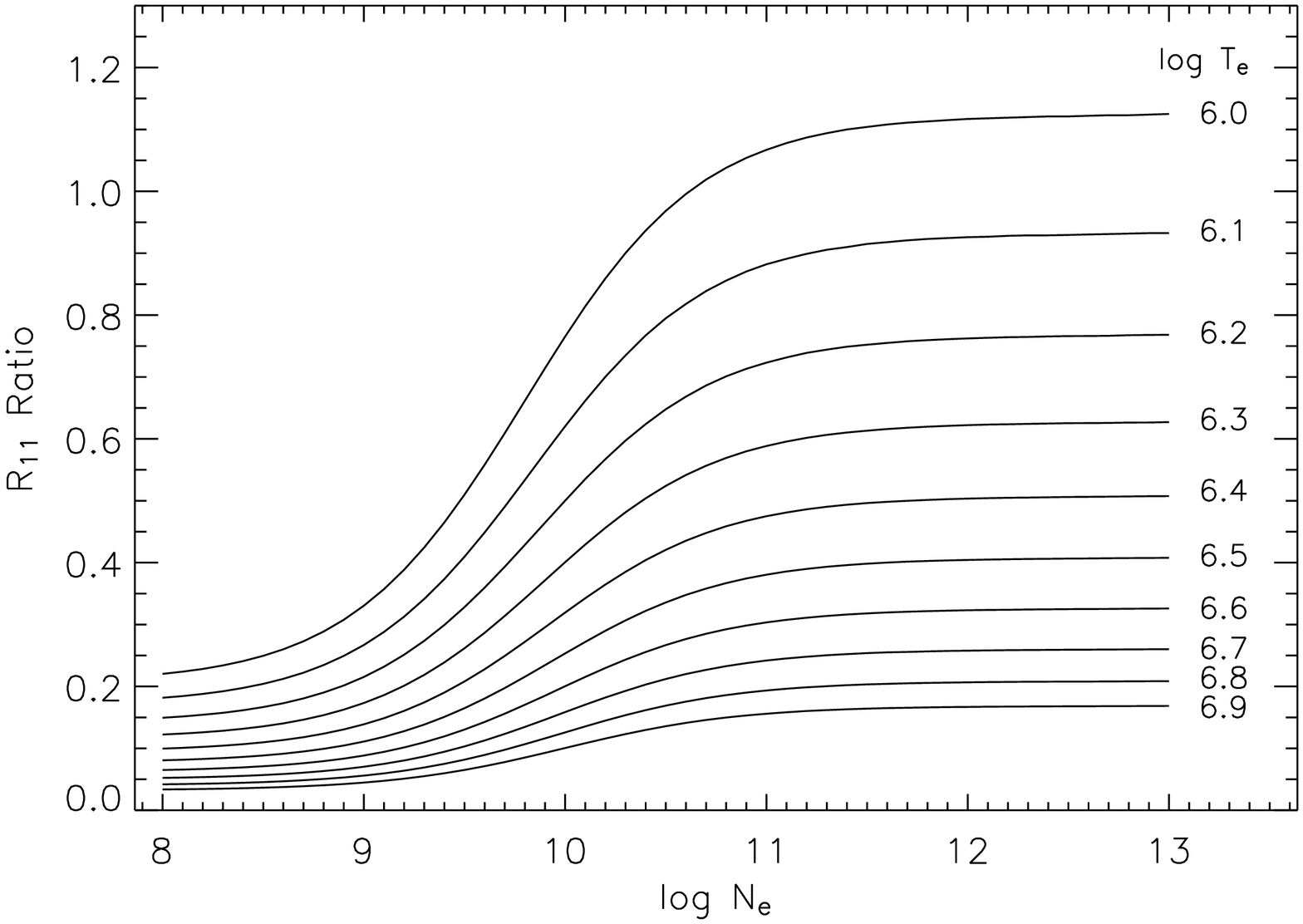}
\caption{
The theoretical \ion{Fe}{15}
emission line intensity ratio
R$_{11}$ = I(82.76 \AA)/I(59.40 \AA),
where I is in photon units,
plotted as a function of logarithmic electron density
(N$_{e}$ in cm$^{-3}$) for several values of logarithmic
electron temperature in the range log T$_{e}$ = 6.0--6.9 (T$_{e}$
in K).  
}
\end{figure}

\begin{figure} 
\includegraphics[scale=1.0,angle=0]{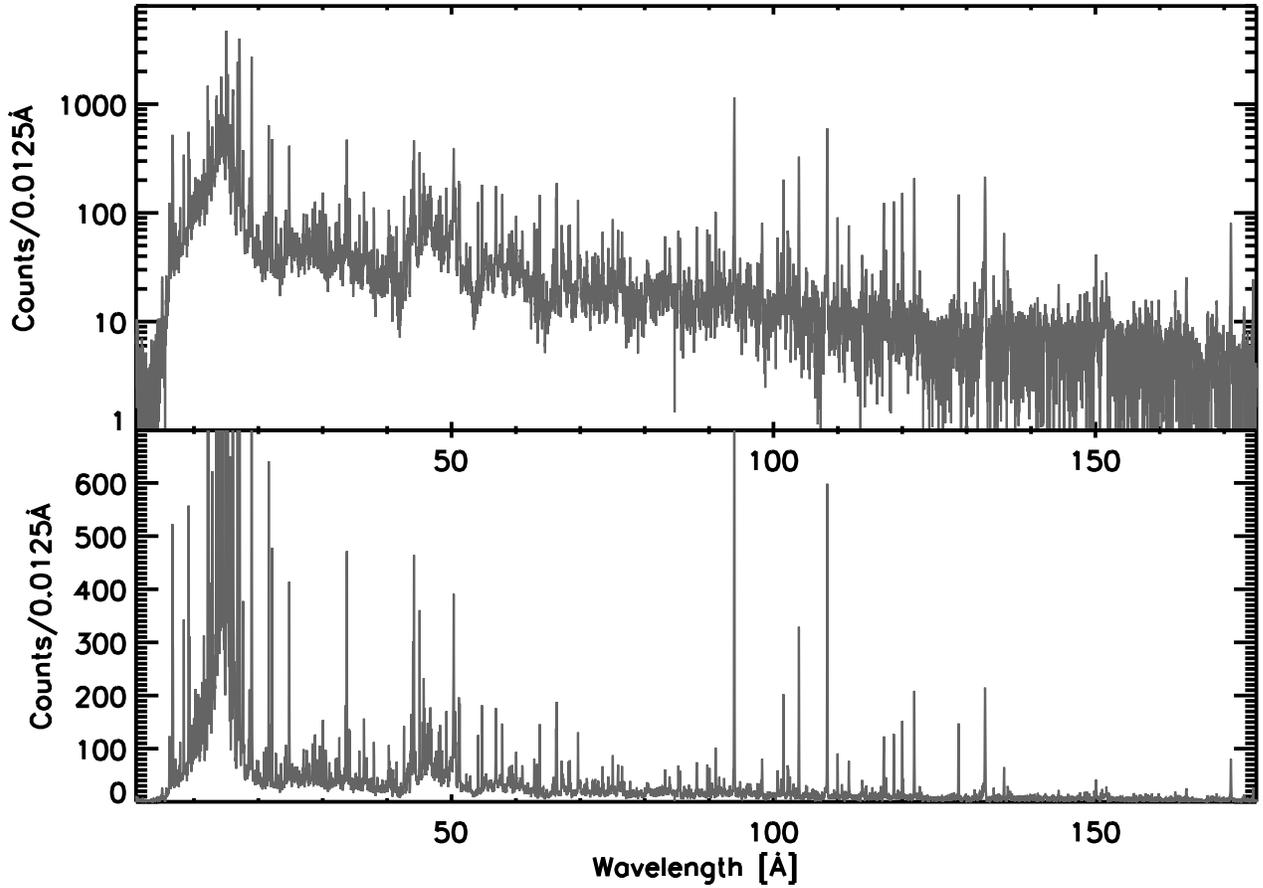}
\caption{
The co-added LETG+HRC-S spectrum of Capella in the 5--175 \AA\ range.
The top panel shows the full dynamic range of the 
spectrum, while the bottom panel is an expanded spectrum to more clearly
show the emission features 
in the 25--175 \AA\ soft X-ray wavelength region.}
\end{figure}

\begin{figure} 
\includegraphics[scale=1.0,angle=0]{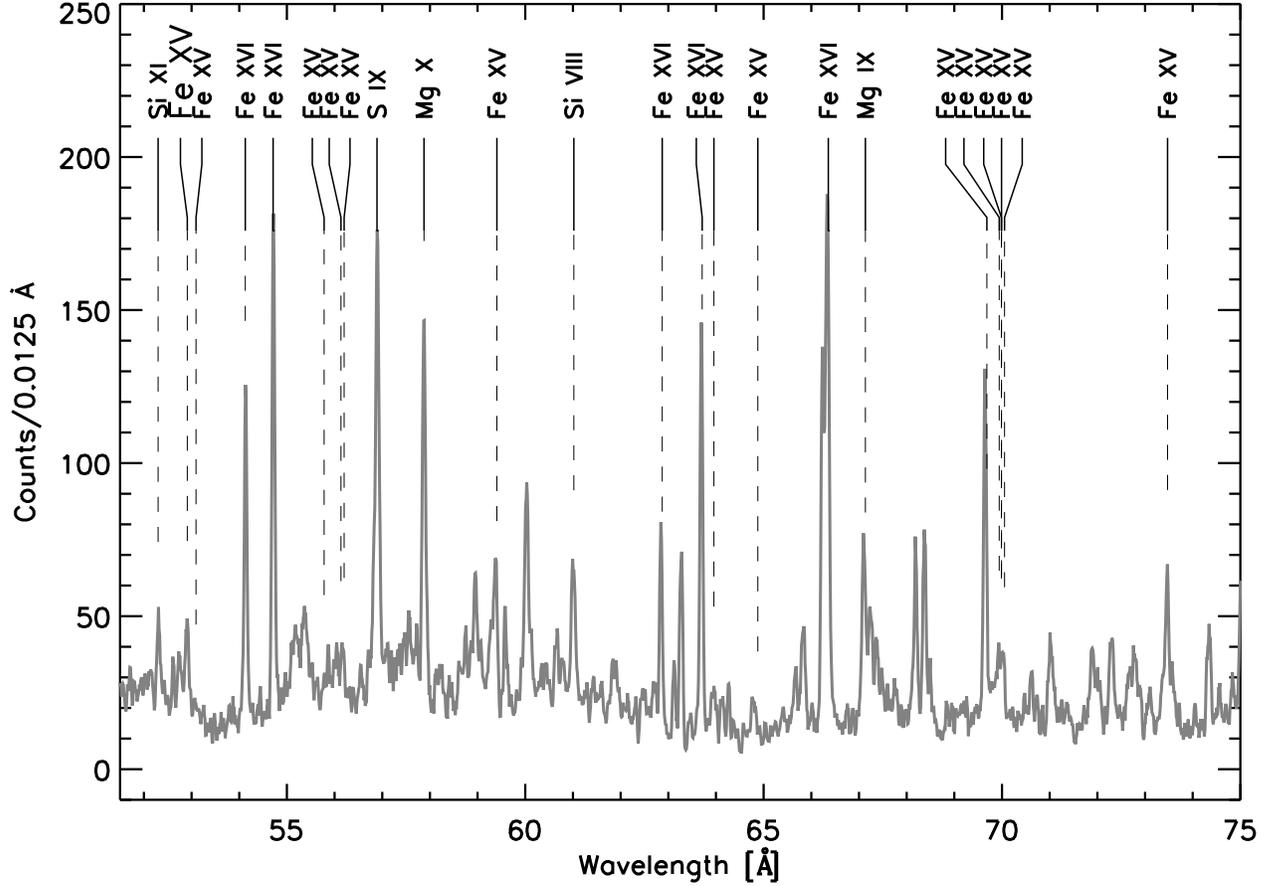}
\caption{
Portion of the co-added
LETG+HRC-S spectrum of Capella in the 50--75 \AA\ range.
Lines of \ion{Fe}{15} analysed in the present paper 
are identified, together with some other
prominent transitions of Fe, Mg, Si and S.  The pseudo-continuum is comprised
of a superposition of weak lines from the {\em n} = 2 levels of abundant
elements, mostly Mg, Si, S and Ar, but also including contributions
from the less abundant elements Na and Al.}
\end{figure}

\begin{figure} [p]
\epsscale{0.4}
\plotone{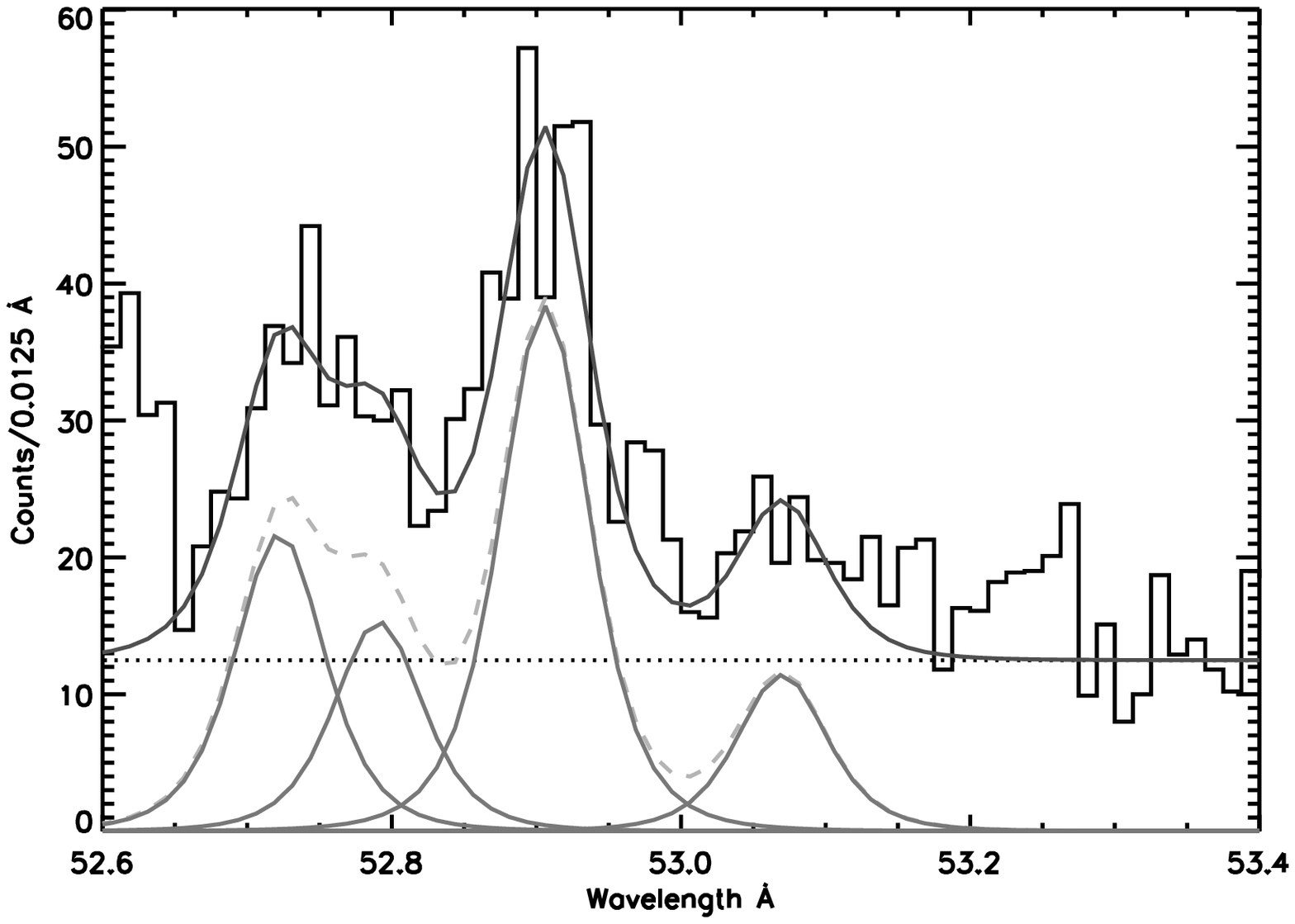}
\plotone{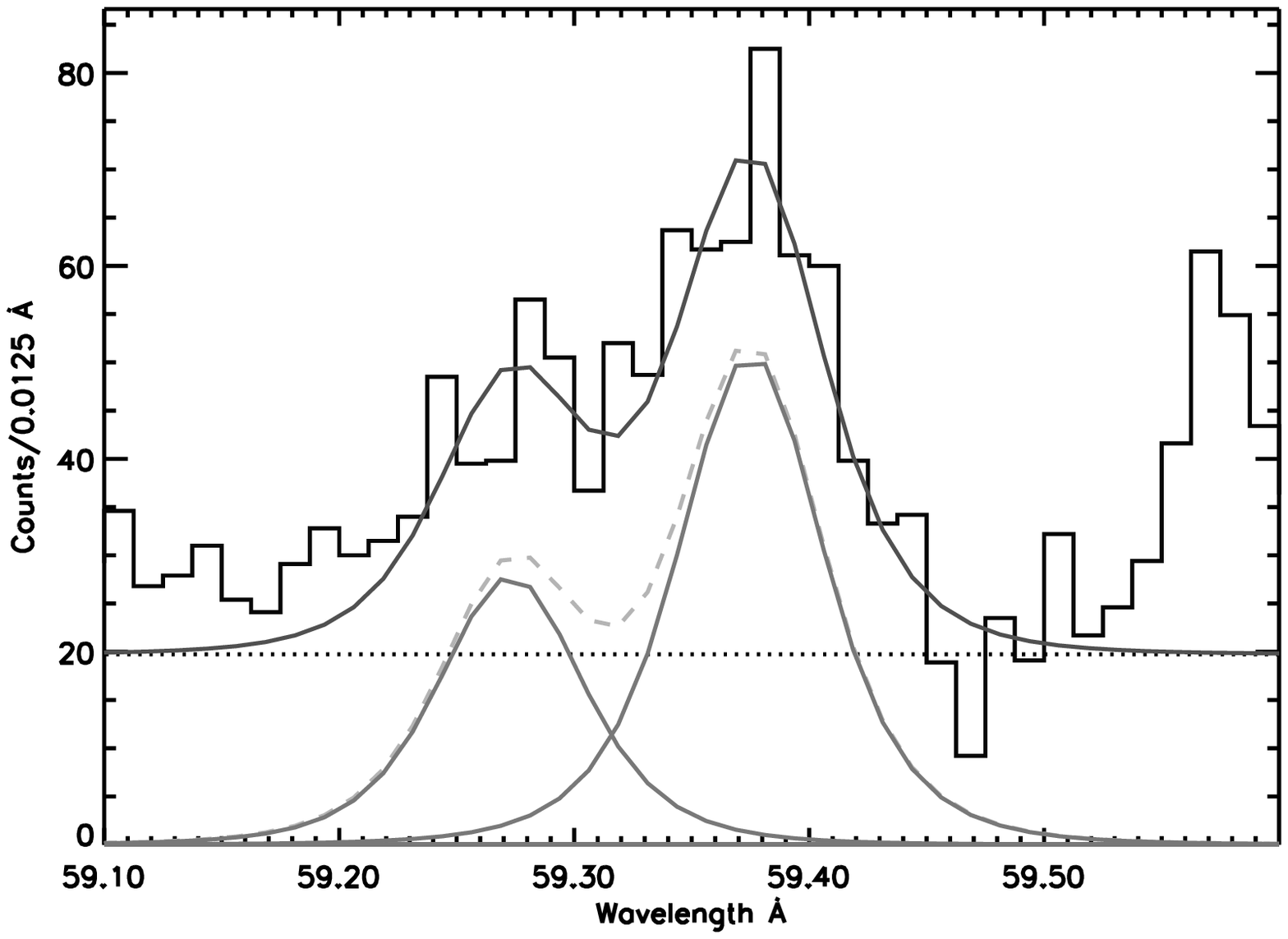}
\plotone{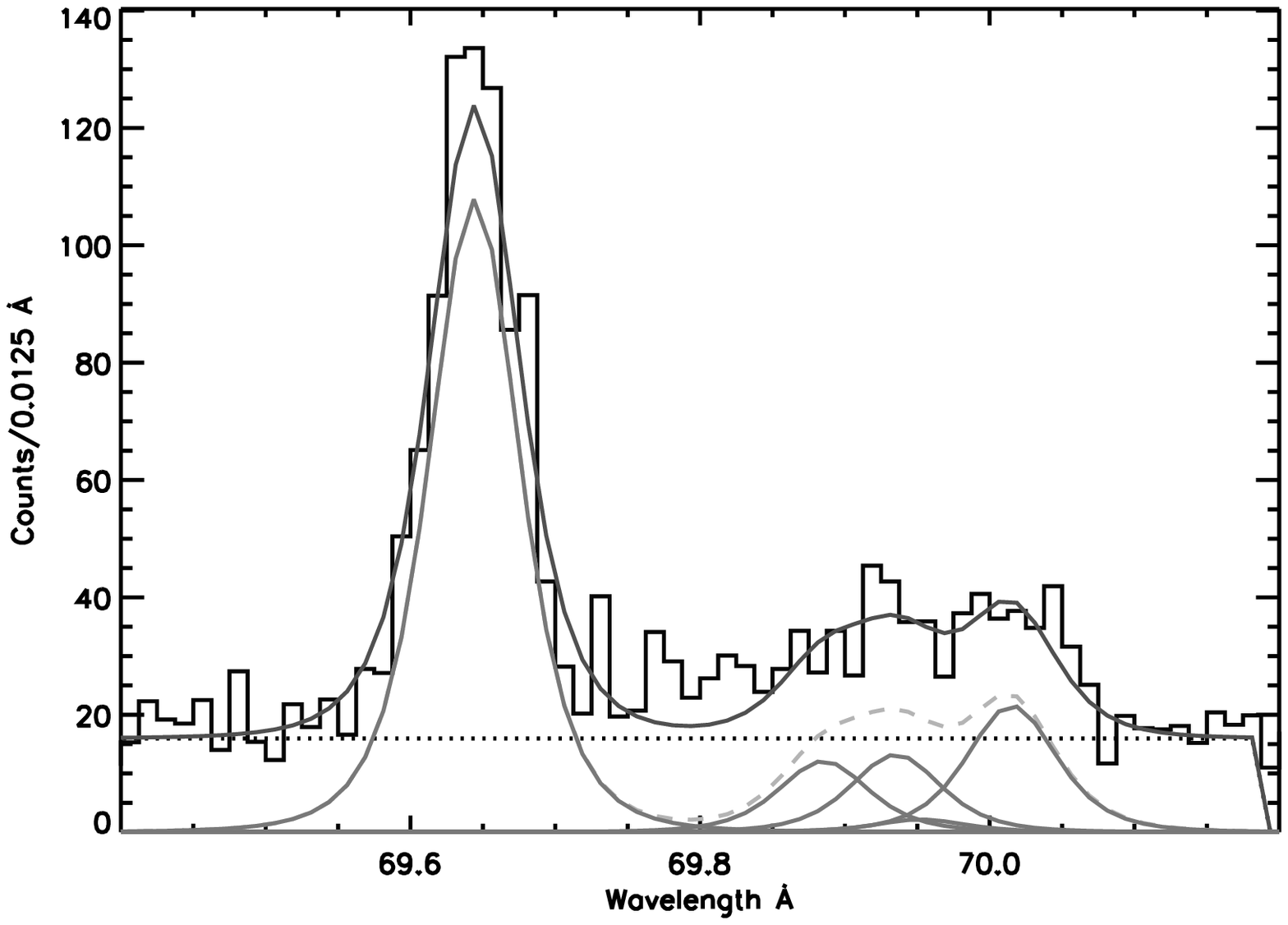}
\plotone{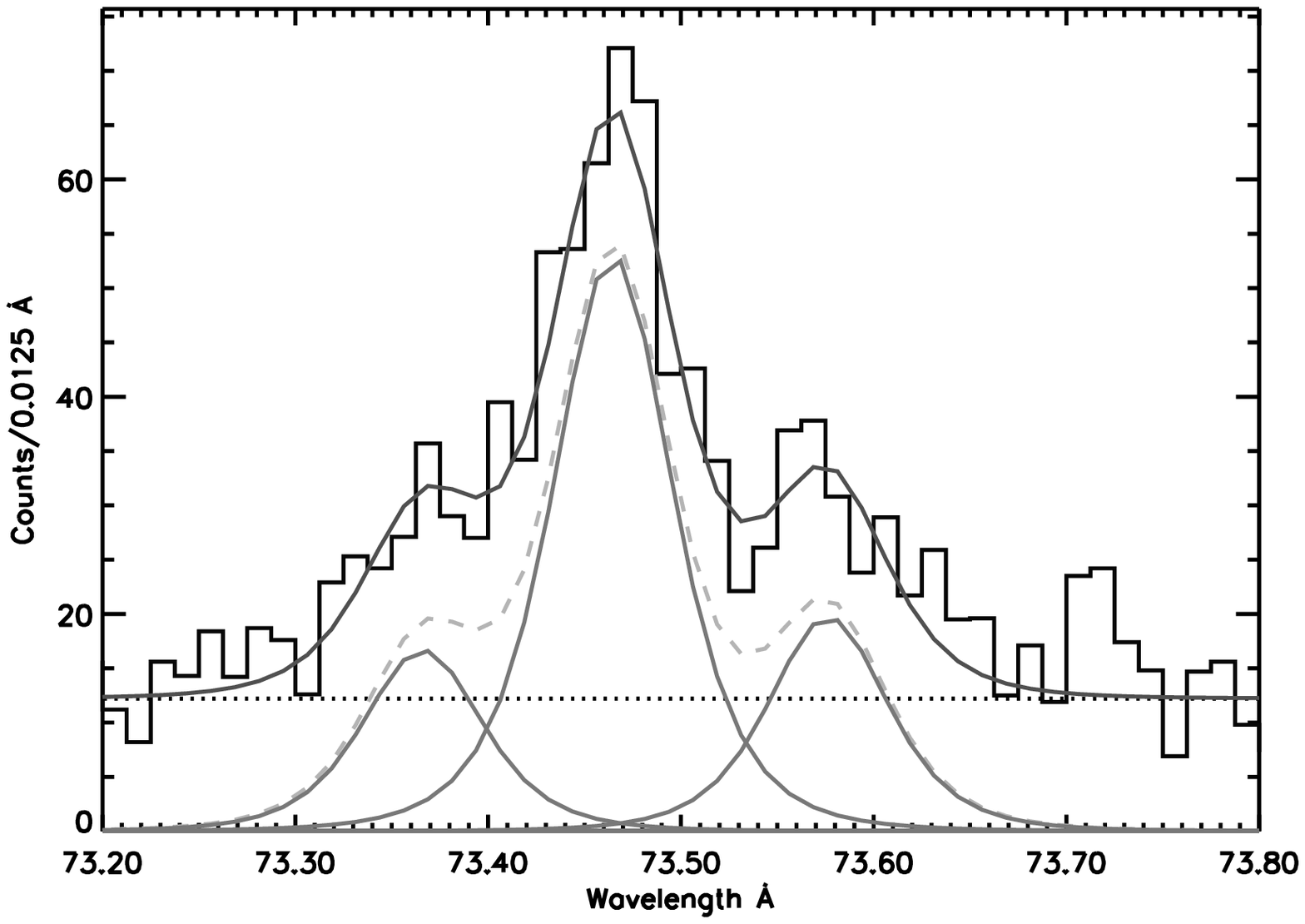}
\caption{Example model fits to the 52.91, 53.11, 59.40, 
69.65, 69.93, 69.98, 70.05 and 73.47 \AA\ 
lines of \ion{Fe}{15} in the {\em Chandra} spectrum of Capella, 
from which the fluxes were
estimated.  Some lines required moderate deblending, which was achieved
by the ad hoc addition of neighbouring lines in the model
fit. 
Adopted pseudo-continua (dotted lines), together with 
individual line components and their sum, are also shown.}
\end{figure}


\clearpage


\begin{deluxetable}{lcc}
\tablecaption{\ion{Fe}{15} transitions in the XSST solar flare spectrum and 
line ratio designations}
\tablehead{
\colhead{Wavelength (\AA)} &
 \colhead{Transition} &
 \colhead{R = I($\lambda$)/I(59.40 \AA)}
}                                                
\startdata
52.91 & 3{\em s}$^{2}$ $^{1}$S$_{0}$--3{\em s}4{\em 
p} $^{1}$P$_{1}$ & R$_{1}$
\\
53.11 & 3{\em s}$^{2}$ $^{1}$S$_{0}$--3{\em s}4{\em 
p} $^{3}$P$_{1}$ & R$_{2}$
\\
55.78 & 3{\em s}3{\em p} $^{3}$P$_{1}$--3{\em s}4{\em 
d} $^{3}$D$_{1,2}$ & R$_{3}$
\\
56.17 & 3{\em s}3{\em p} $^{3}$P$_{2}$--3{\em s}4{\em 
d} $^{3}$D$_{3}$ & R$_{4}$
\\
59.40 & 3{\em s}3{\em p} $^{1}$P$_{1}$--3{\em s}4{\em 
d} $^{1}$D$_{2}$ & \nodata
\\
63.97 & 3{\em p}$^{2}$ $^{1}$D$_{2}$--3{\em s}4{\em 
f} $^{1}$F$_{3}$ & R$_{5}$
\\
66.25 & 3{\em s}3{\em p} $^{3}$P$_{2}$--3{\em s}4{\em 
s} $^{3}$S$_{1}$ & R$_{6}$
\\
69.65 & 3{\em s}3{\em p} $^{1}$P$_{1}$--3{\em s}4{\em 
s} $^{1}$S$_{0}$ & R$_{7}$
\\
69.93 & 3{\em s}3{\em d} $^{3}$D$_{1}$--3{\em s}4{\em 
f} $^{3}$F$_{2}$ & R$_{8}$
\\
69.98 & 3{\em s}3{\em d} $^{3}$D$_{2}$--3{\em s}4{\em 
f} $^{3}$F$_{2,3}$ & R$_{9}$
\\
70.05 & 3{\em s}3{\em d} $^{3}$D$_{3}$--3{\em s}4{\em 
f} $^{3}$F$_{3,4}$ & R$_{10}$
\\
73.47 & 3{\em s}3{\em d} $^{1}$D$_{2}$--3{\em s}4{\em 
f} $^{1}$F$_{3}$ & R$_{12}$
\\
82.76 & 3{\em s}3{\em d} $^{3}$D$_{3}$--3{\em s}4{\em 
p} $^{3}$P$_{2}$ & R$_{11}$
\\
\enddata
\end{deluxetable}

\clearpage


\begin{deluxetable}{lcc}
\tablecaption{\ion{Fe}{15} line ratios in the XSST solar flare spectrum}
\tablehead{
\colhead{Line ratio} &
 \colhead{Observed\tablenotemark{a}} &
 \colhead{Theoretical\tablenotemark{b}}
}                                                
\startdata
R$_{1}$ & 0.78 $\pm$ 0.23 & 0.58 $\pm$ 0.12
\\
R$_{2}$ & 0.36 $\pm$ 0.11 & 0.18 $\pm$ 0.04
\\
R$_{3}$ & 0.33 $\pm$ 0.10 & 0.13 $\pm$ 0.03
\\
R$_{4}$ & 0.54 $\pm$ 0.16 & 0.23 $\pm$ 0.05
\\
R$_{5}$ & 0.35 $\pm$ 0.08 & 0.40 $\pm$ 0.08
\\
R$_{6}$ & 2.6 $\pm$ 0.8 & 0.23 $\pm$ 0.05
\\
R$_{7}$ & 3.0 $\pm$ 0.9 & 3.6 $\pm$ 0.7
\\
R$_{8}$ & 0.28 $\pm$ 0.08 & 0.16 $\pm$ 0.03
\\
R$_{9}$ & 0.33 $\pm$ 0.10 & 0.25 $\pm$ 0.05
\\
R$_{10}$ & 0.73 $\pm$ 0.22 & 0.46 $\pm$ 0.09
\\
R$_{11}$ & 0.75 $\pm$ 0.23 & 0.50 $\pm$ 0.10
\\
R$_{12}$ & 1.4 $\pm$ 0.4 & 1.2 $\pm$ 0.2
\\
\enddata
\tablenotetext{a}{I(59.40 \AA) = 85 photons cm$^{-2}$ s$^{-1}$ arcsec$^{-2}$.}
\tablenotetext{b}{Determined from Figures 1--7 at T$_{e}$ = 10$^{6.3}$ K and
N$_{e}$ = 10$^{10.4}$ cm$^{-3}$.}
\end{deluxetable}

\clearpage


\begin{deluxetable}{lccccc}

\tablecaption{{\em Chandra} LETG+HRC-S observations of Capella employed in the present 
analysis}

\tablehead{\colhead{Obs} & \colhead{Exposure} & \colhead{Start Date} &
\colhead{Start Time} & \colhead{End Date} & \colhead{End Time} \\  
\colhead{(ID)} & \colhead{(ks)} & \colhead{(UT)} & \colhead{(UT)} &
\colhead{(UT)} & \colhead{(UT)} }  

\startdata
58 & 34.11 & 2000-03-08 & 06:30:50 & 2000-03-08 & 16:25:31 \\
62435 & 32.71 & 1999-09-06 & 00:27:21 & 1999-09-06 & 09:49:59 \\
1009 & 26.97 & 2001-02-14 & 11:41:47 & 2001-02-14 & 19:27:44 \\
1248 & 85.23 & 1999-11-09 & 13:28:25 & 1999-11-10 & 13:28:55 \\
1420 & 30.19 & 1999-10-29 & 22:31:27 & 1999-10-30 & 07:29:02 \\
3675 & 27.16 & 2003-09-28 & 04:23:10 & 2003-09-28 & 12:21:25 \\
\enddata
\label{t:obs}
\end{deluxetable}

\clearpage


\begin{deluxetable}{lcc}
\tablecaption{\ion{Fe}{15} line ratios in the {\em Chandra} observations of 
Capella}
\tablehead{
\colhead{Line ratio} &
 \colhead{Observed\tablenotemark{a}} &
 \colhead{Theoretical\tablenotemark{b}}
}                                                
\startdata
R$_{1}$ & 0.55 $\pm$ 0.06 & 0.58 $\pm$ 0.12
\\
R$_{2}$ & 0.16 $\pm$ 0.03 & 0.18 $\pm$ 0.04
\\
R$_{4}$ & 0.26 $\pm$ 0.06 & 0.21 $\pm$ 0.04
\\
R$_{5}$ & 0.39 $\pm$ 0.06 & 0.40 $\pm$ 0.08
\\
R$_{7}$ & 2.9 $\pm$ 0.2 & 3.6 $\pm$ 0.7
\\
R$_{8}$ & 0.28 $\pm$ 0.06 & 0.15 $\pm$ 0.03
\\
R$_{9}$ & 0.35 $\pm$ 0.07 & 0.24 $\pm$ 0.05
\\
R$_{10}$ & 0.49 $\pm$ 0.06 & 0.44 $\pm$ 0.09
\\
R$_{12}$ & 1.3 $\pm$ 0.1 & 1.2 $\pm$ 0.2
\\
\enddata
\tablenotetext{a}{I(59.40 \AA) = (9.7 $\pm$ 0.7) $\times$ 10$^{-5}$ photons 
cm$^{-2}$ s$^{-1}$.}
\tablenotetext{b}{Determined from Figures 1--7 at T$_{e}$ = 10$^{6.3}$ K and
N$_{e}$ = 10$^{10.2}$ cm$^{-3}$.}
\end{deluxetable}

\clearpage


\begin{deluxetable}{lcl}
\tablecaption{Wavelengths of ad hoc line components used in the Capella 
analysis}
\tablehead{
\colhead{Fe XV $\lambda$ (\AA )} &
\colhead{Blend $\lambda$ (\AA )} &
\colhead{Candidate identification}
}

\startdata

52.91 &
    52.72  &  \ion{Fe}{17}? \ion{Ni}{18}? \\
    & 52.79    &  \ion{Fe}{18}? \ion{S}{9}? \\

55.78 &  
    55.87  &   \ion{Si}{9}? \\

56.17 &  
    56.06     &  \ion{Ne}{9}? \ion{S}{9}? \ion{Fe}{17}? \\

59.40 &  
    59.27    &  \ion{Fe}{17}?  \\

63.97 &  
    64.00    &  \ion{O}{8} + \ion{Fe}{18} 4th order \\

73.47 &  
    73.37    &  \\
    &     73.58    &  \ion{Ne}{8}? \\

\enddata
\end{deluxetable}

\end{document}